\documentclass[aps,pre,twocolumn]{revtex4-2}

\usepackage{xcolor,amsmath,amssymb}
\usepackage{graphicx}
\usepackage{multirow}

\definecolor{darkblue}{rgb}{0,0,0.6}
\definecolor{darkred}{rgb}{0.6,0,0}
\usepackage[colorlinks=true,urlcolor=darkblue,citecolor=darkblue, linkcolor=darkred,hyperfootnotes=false]{hyperref}

\definecolor{pinegreen}{rgb}{0.0, 0.47, 0.44}

\newcommand{\wt}{\widetilde}
\newcommand{\ga}{0}

\newcommand{\mcR}{\mathcal{R}}
\newcommand{\mcS}{\mathcal{S}}

\newcommand{\dd}{\text{d}}

\newcommand{\ind}[1]{_{\mathrm{#1}}}

\newcommand{\ed}{\mathrm{e}}

\newcommand{\id}{\mathrm{i}}

\newcommand{\kk}{\boldsymbol{k}}
\newcommand{\mcO}{\mathcal{O}}
\newcommand{\PP}{\boldsymbol{P}}
\newcommand{\qq}{\boldsymbol{q}}

\newcommand{\xx}{\boldsymbol{x}}
\newcommand{\XX}{\boldsymbol{X}}

\newcommand{\YY}{\boldsymbol{Y}}

\newcommand{\eeta}{\boldsymbol{\eta}}

\newcommand{\psitt}{\psi_2^{(2)}}
\newcommand{\psitf}{\psi_2^{(4)}}

\begin{document}

\title{Non-Gaussian fluctuations of a probe coupled to a Gaussian field}

\author{Vincent D\'emery}
\address{Gulliver, CNRS, ESPCI Paris PSL, 10 rue Vauquelin, 75005 Paris, France}
\address{Univ Lyon, ENS de Lyon, CNRS, Laboratoire de Physique, F-69342 Lyon, France}

\author{Andrea Gambassi}
\address{SISSA --- International School for Advanced Studies and INFN, via Bonomea 265, 34136 Trieste, Italy}

\date{ \today }

\begin{abstract}
The motion of a colloidal probe in a complex fluid, such as a micellar solution, is usually described by the generalized Langevin equation, which is linear.
However, recent numerical simulations and experiments have shown that this linear model fails when the probe is confined, and that the intrinsic dynamics of the probe is actually non-linear.
Noting that the kurtosis of the displacement of the probe may reveal the non-linearity of its dynamics also in the absence confinement, we compute it for a probe coupled to a Gaussian field and possibly trapped by a harmonic potential.
We show that the excess kurtosis increases from zero at short times, reaches a maximum, and then decays algebraically at long times, with an exponent which depends on the spatial dimensionality and on the features and correlations of the dynamics of the field.
Our analytical predictions are confirmed by numerical simulations of the stochastic dynamics of the probe and the field where the latter is represented by a finite number of modes.
\end{abstract}

\maketitle

\section{Introduction}
\label{sec:intro}

Complex fluids, such as polymer solutions or colloidal suspensions, are usually described by their frequency-dependent complex shear modulus, which characterizes the linear response of the fluid~\cite{Ferry1980Viscoelastic}. 
The motion of a colloidal probe immersed in such fluids then follows a linear generalized Langevin equation (GLE), with a memory kernel which is directly related to the complex modulus of the fluid~\cite{Mazur1978,Mason1995}.
With such a GLE, the temporal correlations of the velocity of the probe can be expressed as a function of the memory kernel~\cite{MedinaNoyola1987}.
In microrheology, this relation is inverted in order to infer the complex modulus of the fluid from the observation of the motion of the probe~\cite{Mason1995}.
This technique is now widely used to probe complex fluids~\cite{Mizuno2001Electrophoretic, Mizuno2007Nonequilibrium, Wirtz2009}.

In a micellar solution, the GLE has recently been successful in describing the transitions of a probe in a double well potential~\cite{Ginot2022Barrier} and the recoil dynamics when the external trap which drags the probe through the fluid is suddenly switched off \cite{Ginot2022Recoil}.
This description 
is valid as long as the fluid remains within 
the linear response regime.
If the probe is driven sufficiently fast, as in active microrheology~\cite{Meyer2006,Wilson2011b}, the response of the fluid is non-linear and thus the dynamics of the probe cannot be captured by the GLE, which is linear~\cite{GomezSolano2014,GomezSolano2015}.
Similarly, recent molecular dynamics simulations and experiments have shown that the GLE encounters limitations even in describing the motion of a probe in a static trap at equilibrium~\cite{Daldrop2017, Muller2020}.
In these examples, the probe is held in a harmonic trap and an effective memory kernel is infereed from the temporal correlations of the position of the probe by assuming that the motion of the probe is described by the GLE. 
The resulting effective memory kernel, however, turns out to depend on the stiffness of the trap, meaning that it does not describe solely the coupling between the probe and the solvent.
Accordingly, while on the one hand the linear GLE has been a successful model of the effective dynamics of probe particles, on the other, the accumulating evidence that it cannot fully capture it calls for the introduction of more general, non-linear models for the probe dynamics. 

A stochastic Prandtl-Tomlinson model, where the probe is coupled to a virtual ``bath particle'' via a sinusoidal potential, has been proposed to describe the interaction between the probe and a micellar solution~\cite{Jain2021Two, Jain2021Microrheology}, but the parameters fitted to the correlations also depend on the parameters of the confining potential, e.g., on its stiffness~\cite{Muller2020}.
Recently, it has been shown that coupling the probe to a scalar Gaussian field also reproduces qualitatively the stiffness dependence of the effective memory kernel~\cite{Basu2022Dynamics}.
In this model, the field can be integrated out, resulting in a non-Markovian dynamics for the probe, with a non-linear memory kernel and a colored noise, which allowed the perturbative analytical calculation of the first and second moments of the probe displacement~\cite{Demery2011Perturbative,Demery2014c,Demery2019}.
This effective non-Markovian dynamics turned out to affect the relaxation of the average position of the probe after being released from outside the center of a harmonic trap~\cite{Venturelli2022Nonequilibrium}, to determine the synchronization of the motion of two probe particles interacting with the same field~\cite{Venturelli2022Inducing}, the spatial distribution of a single probe when the field is confined within a film~\cite{Venturelli2022Tracer}, or it might even give rise to oscillations of the average position of the probe when the harmonic trap is dragged at a constant speed~~\cite{Venturelli2023Memory-induced}.
In this case, as a function of the dragging speed, the effective friction shows a rich behaviour beyond Stoke's law~\cite{Venturelli2023Stochastic}.
While the emergence of some of the phenomena investigated in these studies hinges upon the non-linear and the non-Gaussian nature of the statistical fluctuations in the effective equation of motion of the probe, it would be desirable to identify and focus on more direct markers of these two features.

The kurtosis of the displacement emerges, in this respect, as a 
natural observable to characterize the non-linearity of the equations of motion~\cite{Metzler2019Brownian}.
Indeed, in the presence of Gaussian noise in linear equations of motion, the displacement of the probe between two arbitrary times is a Gaussian random variable, and therefore its kurtosis takes the value 3.
An advantage of focussing on the kurtosis is that there is no need to determine the effective memory kernel and its dependence on the parameters of the system in order to demonstrate a non-linear behavior.
Another advantage is that the kurtosis can be measured for a probe held in a harmonic trap as well as when it is freely diffusing.
Because of these advantages, the kurtosis, and more generally the deviations of the statistics of the probe displacements  from a Gaussian distribution, have been widely used in order to characterize the dynamics of particles in complex systems, such as entangled actin networks~\cite{Wang2009} or baths of swimmers~\cite{Ortlieb2019, Lagarde2020}.

Here, we compute the kurtosis of the spatial displacement in a time $t$ of a probe particle coupled to a Gaussian field and held in a harmonic trap. Both the probe particle and the field are in contact with an equilibrium thermal bath.
Our calculation is perturbative in the coupling between the probe and the field, 
and relies on the path integral method introduced in Ref.~\cite{Demery2011Perturbative}.
This method has been used previously in order to compute first and second orders moments of the displacement~\cite{Demery2011Perturbative,Demery2014c,Demery2019}; here we extend this approach to the determination of a fourth-order moment. 
Based on this analysis, we find that the excess kurtosis --- the difference between its actual value and that for Gaussian fluctuations --- grows at short times as $t^2$, reaches a maximum, and then it decays algebraically with an exponent which we determine as a function of the properties of the field, the dimensionality $d$ of space, and of the presence or absence of the trap.
Our predictions are confirmed by numerical simulations of the stochastic dynamics in $d=1$, where the Gaussian field is represented by 10 independent modes.

This article is organized as follows:
In Sec.~\ref{sec:model}, we introduce the model and the observables that we compute.
In Sec.~\ref{sec:pi_calc}, we calculate the second- and fourth-order moments of the displacement of the probe using the path-integral formalism.
In Sec.~\ref{sec:simulations}, we describe the simulations and compare their results with our theoretical predictions.
In Sec.~\ref{sec:asymptotic}, we analyse the short- and long-time asymptotic behavior of the second- and fourth-order moments of the displacement.
We present our conclusions and perspectives in Sec.~\ref{sec:conclusion}.

\section{The model}\label{sec:model}

\subsection{Definition}\label{}

We consider a probe with position $\XX(t)\in\mathbb{R}^d$ held in a harmonic trap with stiffness $\kappa$ and coupled to a Gaussian field $\phi(\xx,t)$~\cite{Demery2010,Demery2010a, Demery2011Perturbative, Demery2019,Basu2022Dynamics,Venturelli2022Nonequilibrium,Venturelli2022Inducing,Venturelli2022Tracer,Venturelli2023Memory-induced,Venturelli2023Stochastic}. 
The energy of the system is given by
\begin{equation}\label{eq:energy}
H[\XX,\phi]=\frac{\kappa}{2}\XX^2+\frac{1}{2}\int\!\!\dd\xx\, \phi(\xx)A\phi(\xx) - K\phi(\XX).
\end{equation}
The operator $A$ describes the energy of the field $\phi(\xx,t)$ and $K$ determine its linear coupling to the probe.

We assume an overdamped Langevin dynamics for the system composed by the probe and the field. In particular, 
the dynamics of the field reads
\begin{align}
\dot \phi(\xx,t) & = -R \frac{\delta H}{\delta\phi(\xx,t)} + \xi(\xx,t)\\
& = -RA\phi(\xx,t) + RK\delta(\xx-\XX(t))+\xi(\xx,t), \label{eq:dyn_field}
\end{align}
where $R$ is the mobility operator and $\xi(\xx,t)$ a Gaussian white noise with correlation function
\begin{equation}
\left\langle \xi(\xx,t)\xi(\xx',t') \right\rangle = 2TR(\xx-\xx')\delta(t-t'),
\label{eq:corr-xi}
\end{equation}
where $T$ is the thermal energy. As anticipated in the introduction, the dynamics of the field is determined, inter alia, by the coupling to an \emph{equilibrium} thermal bath and therefore the dynamics prescribed by Eqs.~\eqref{eq:dyn_field} and \eqref{eq:corr-xi} satisfy the fluctuation-dissipation relation. 

Concrete examples of physical relevance which have been considered in the literature so far assume, e.g.,
a Gaussian model for the field with 
\begin{equation}
\label{eq:A-Gaux}
A = - \nabla^2 + m^2, 
\end{equation}
in which the fluctuations of the field in equilibrium are correlated across a typical distance set by the correlation length $\xi = m^{-1}$. The coupling between the field and the probe, instead, is assumed to be 
\begin{equation}
K\phi(\XX) \equiv \int \dd\xx \, K(\xx-\XX)\phi(\xx),
\end{equation}
where the integral kernel $K(\xx)$ is an isotropic function which vanishes rapidly upon increasing $|\xx|$ beyond a certain value $a$, representing the ``radius'' of the probe. In Sec.~\ref{sec:simulations} we shall consider explicitly the case $K(\xx) = \delta(\xx)$, corresponding to a point probe and in Sec.~\ref{sec:asymptotic} the case of a Gaussian dependence of $K$ on $|\xx|$. 
Concerning the dynamics of the field, instead, natural choices of the operator $R$ correspond to a purely relaxational dynamics of the field (the so-called model A of Ref.~\cite{Hohenberg1977}) with constant $R$ or to a dynamics with local conservation of the field (model B of Ref.~\cite{Hohenberg1977}) with $R \propto - \nabla^2$, which are considered in Secs.~\ref{sec:simulations} and \ref{sec:asymptotic}.

Similarly to the case of the field, the overdamped Langevin dynamics of the probe is given by
\begin{align}
\dot\XX(t) & = -\gamma^{-1}\nabla_{\XX} H + \eeta(t) \nonumber\\
& = -\omega_0\XX(t)+\gamma^{-1}\nabla K\phi(\XX(t),t)+\eeta(t), \label{eq:dyn_probe}
\end{align}
where $\gamma$ is the friction coefficient of the probe due to the solvent, $\omega_0=\kappa/\gamma$ is the relaxation rate of the position of the probe in the harmonic trap and $\eeta(t)$ is a Gaussian white noise with correlation function
\begin{equation}
\left\langle \eta_\mu(t)\eta_\nu(t') \right\rangle = 2D_0\delta_{\mu\nu}\delta(t-t'),
\label{eq:corr-noise-X}
\end{equation}
where $\mu$, $\nu\in\{1,\ldots,d\}$ indicate the various spatial components of a vector.
In the expression above we have introduced the diffusion coefficient $D_0=T/\gamma$ of the probe for $K=0$, i.e., in the absence of the coupling to the field, which we refer to as the \emph{bare} diffusion coefficient, denoted by the subscript $0$. 
As in the case of $\phi(\xx,t)$, the dynamics of the particle satisfies the fluctuation-dissipation relation and, as a consequence, the distribution function of the field and the particle in equilibrium is determined by the Boltzmann factor  $\propto {\rm e}^{-H/T}$ with the Hamiltonian $H$ in Eq.~\eqref{eq:energy} and the thermal energy $T$.

\subsection{Cumulant generating function}\label{subsec:cgf}

As anticipated in Sec.~\ref{sec:intro}, we consider here the particle trajectory $\XX(t)$ and focus on the statistics of its displacements. In particular, we assume that at the initial time $t_0$, the particle is at position $\XX(t_0)$ and, correspondingly,  the field has a certain configuration $\phi(\xx,t_0)$ (alternatively, these position and configuration may be drawn from some probability distributions).  In the limit $t_0\to - \infty$ considered here, the initial conditions do not affect the dynamics of the system at time $t\ge 0$ (both in the presence or in the absence of the trap), which becomes stationary, and thus the displacement within a time interval $t$ can be simply expressed as $\YY(t)=\XX(t)-\XX(0)$.
In order to compute the moments of  $\YY(t)$ we will use the cumulant generating function
\begin{equation}
\psi(\qq,t)= \ln \left\langle \ed^{\id\qq\cdot\YY(t)} \right\rangle.
\label{eq:def-cgf}
\end{equation}
Expanding this function in powers of $\qq$ gives access to the cumulants of the displacement $\YY$.
For an isotropic system, $\psi(\qq,t)$ depends only on the norm $q = |\qq|$, the odd cumulants vanish, and the expansion reads
\begin{equation}
\psi(\qq,t) = -\frac{1}{2}\psi^{(2)}(t)q^2+\frac{1}{4!}\psi^{(4)}(t)q^4 + \mcO(q^6).
\label{eq:psi-dec}
\end{equation}
In terms of the coefficients $\psi^{(2)}(t)$ and $\psi^{(4)}(t)$ of this expansion, 
the cumulants of the components of $\YY$ are given by
\begin{align}
\langle Y_\mu(t)Y_\nu(t) \rangle_c &= \psi^{(2)}(t)\delta_{\mu\nu},\label{eq:def-psi2}\\
\langle Y_\mu(t)Y_\nu(t)Y_\sigma(t)Y_\tau(t) \rangle_c &= \psi^{(4)}(t)\nonumber\\&\times
\frac{\delta_{\mu\nu}\delta_{\sigma\tau}+\delta_{\mu\sigma}\delta_{\nu\tau}+\delta_{\mu\tau}\delta_{\nu\sigma}}{3} .
\end{align}
The fourth cumulant of the displacement along the direction 1 is
\begin{equation}
\langle Y_1^4(t) \rangle_c = \langle Y^4_1(t) \rangle-3\langle Y^2_1(t) \rangle^2 = \psi^{(4)}(t),
\end{equation}
and it vanishes if $Y_1$ has a Gaussian distribution.
The excess kurtosis is therefore defined as 
\begin{equation}
\gamma(t) = \frac{\langle Y^4_1(t) \rangle_c}{\langle Y^2_1(t) \rangle^2}
=\frac{\psi^{(4)}(t)}{\left[ \psi^{(2)}(t)\right]^2}.
\label{eq:def-gamma}
\end{equation}
Without coupling to the field, i.e., for $K = 0$,  $\YY(t)$ is Gaussian with variance
\begin{equation}
\langle Y_\mu(t)Y_\nu(t) \rangle_\ga = \sigma_0^2(t)\delta_{\mu\nu},
\end{equation}
where we introduced
\begin{equation}
\sigma_0^2(t) = \frac{2D_0}{\omega_0}\left(1-\ed^{-\omega_0 |t|}\right) =  \frac{2T}{\kappa}\left(1-\ed^{-\omega_0 |t|}\right),
\label{eq:def-sig0}
\end{equation}
and used the definitions of $\omega_0$  and $D_0$ reported after Eqs.~\eqref{eq:dyn_probe} and  \eqref{eq:corr-noise-X}, respectively.
Accordingly, the bare cumulant generating function is given by
\begin{equation}
\label{eq:psiG}
\psi_0(\qq,t)= - \frac{1}{2}\sigma_0^2(t) q^2.
\end{equation}
This cumulant generating function is quadratic in $q$, meaning that $\psi_0^{(4)}(t)= 0$ (see the decomposition in Eq.~\eqref{eq:psi-dec} for the case of $\psi_0(\qq,t)$)  and therefore the excess kurtosis $\gamma(t)$ as well as all the higher-order cumulants vanish at all times.

Here we compute perturbatively the correction to the cumulant generating function $\psi(\qq,t)$ due to the probe-field coupling $K$.
At the lowest non-trivial order in such a coupling, expanding this correction in $q$ gives us access to the correction to the mean-square displacement and to the excess kurtosis.

\section{Path-integral calculation}
\label{sec:pi_calc}

\subsection{Path-integral formalism and perturbative correction}\label{}

The dynamics of the field in Eq.~\eqref{eq:dyn_field} is linear and can be integrated, leading to an effective dynamics of the probe with a non-linear memory~\cite{Demery2011Perturbative,Basu2022Dynamics}.
Following Refs.~\cite{Demery2011Perturbative,Demery2019}, we use a path-integral representation of the effective dynamics of the probe in order to compute perturbatively the correction to the cumulant generating function $\psi(\qq,t)$. 
The perturbative expansion is done in terms of increasing powers of the probe-field coupling $K$ (see Eq.~\eqref{eq:energy}).

The correction to the observable $\ed^{\psi(\qq,t)} = \left\langle \ed^{\id\qq\cdot\YY(t)} \right\rangle$  (see Eq.~\eqref{eq:def-cgf}) is calculated to be
\begin{multline}\label{eq:exp_sint}
\left\langle \ed^{\id\qq\cdot\YY(t)} \right\rangle =  \left\langle \ed^{\id\qq\cdot\YY(t)} \right\rangle_0 - \left\langle \ed^{\id\qq\cdot\YY(t)} S\ind{int}\right\rangle_0 +{\cal O}(K^4)\\
= \ed^{- q^2\sigma_0^2(t)/2}- \left\langle \ed^{\id\qq\cdot\YY(t)} S\ind{int}\right\rangle_0 +{\cal O}(K^4),
\end{multline}
where $S\ind{int}$ is the interaction part of the Janssen-De Dominicis action.
This interaction can be written as (see, e.g., Sec. 4.1 in Ref.~\cite{Tauber2014})
\begin{multline}
\label{eq:Sint-def}
S\ind{int}[\XX,\PP] = \int\!\dd t\, \dd t' P_\mu(t) \left[\id F_\mu(\XX(t)-\XX(t'),t-t')\right.\\\left.+TG_{\mu\nu}(\XX(t)-\XX(t'),t-t') P_\nu(t')\right] \theta(t-t'),
\end{multline}
where $\PP(t)$ is the so-called response field, $\theta(t)$ is the Heaviside function, and $F_\mu(\xx,t)$ and $G_{\mu\nu}(\xx,t)$ are given by
\begin{align}
F_\mu(\XX,t) & = \frac{1}{\gamma}\int \frac{\dd\kk}{(2\pi)^d}\id k_\mu\wt R\wt K^2\ed^{-\wt R\wt At+\id\kk\cdot\XX}, \label{eq:F-def}\\
G_{\mu\nu}(\XX,t)&=\frac{1}{\gamma^2}\int \frac{\dd\kk}{(2\pi)^d} k_\mu k_\nu \frac{\wt K^2}{\wt A}\ed^{-\wt R\wt A|t|+\id\kk\cdot\XX}.\label{eq:G-def}
\end{align}
These expressions involve the Fourier transforms $\wt R(\kk)$, $\wt A(\kk)$ and $\wt K(\kk)$ of the operators $R$, $A$, and $K$, respectively; the dependence on $\kk$ of these transforms  is omitted to lighten the notations.
In addition, below we consider the case of an isotropic system, such that these operators depend on $\kk$ only via its modulus $k=|\kk|$.
Note that the interaction action $S\ind{int}$ is quadratic in the probe-field interaction $K$ and this is the order in perturbation theory which we focus on below.

The averages $\langle \cdots \rangle_\ga$ in Eq.~\eqref{eq:exp_sint} are performed for the unperturbed Gaussian fields $\XX(t)$ and $\PP(t)$, which have zero mean and correlations
\begin{align}
\langle P_\mu(t) P_\nu(t') \rangle_0 & = 0,\\
\langle X_\mu(t) P_\nu(t') \rangle_0 & = \id\ed^{-\omega_0(t-t')}\theta(t-t')\delta_{\mu\nu} \equiv \id\mcR(t-t')\delta_{\mu\nu},\\
\langle X_\mu(t) X_\nu(t') \rangle_0 & = \frac{D_0}{\omega_0} \ed^{-\omega_0 |t-t'|}\delta_{\mu\nu}
=\frac{T}{\kappa}\mcR(|t-t'|)\delta_{\mu\nu},
\end{align}
where we have defined the response function 
\begin{equation}
\mcR(t)=\ed^{-\omega_0 t}\theta(t),
\label{eq:def-R}
\end{equation}
and used the definitions of $\omega_0$  and $D_0$ reported after Eqs.~\eqref{eq:dyn_probe} and  \eqref{eq:corr-noise-X}, respectively.
From these expressions, one can easily calculate the 
mean-square displacement,
\begin{multline}
\left\langle [\XX(t)-\XX(t')]_\mu [\XX(t)-\XX(t')]_\nu \right\rangle_0\\ = \frac{2T}{\kappa}\left[1-\mcR(|t-t'|) \right] \delta_{\mu\nu}
\equiv 
\sigma_0^2(t-t')\delta_{\mu\nu}.
\end{multline}
The cumulant generating function following from Eq.~\eqref{eq:exp_sint} is, at order $K^2$,
\begin{align}
\psi(\qq,t) &= -\frac{q^2}{2} \sigma_0^2(t) - 
\ed^{\frac{q^2}{2}\sigma_0^2(t)}
\left\langle \ed^{\id\qq\cdot\YY(t)} S\ind{int}\right\rangle_0\\
&=\psi_0(\qq,t)+\psi_2(\qq,t), \label{eq:psi-calc-1}
\end{align}
where $\psi_\ga$ is the Gaussian contribution of ${\cal O}(K^0)$  given in Eq.~\eqref{eq:psiG} and $\psi_2$ is due to the coupling to the field and is of 
${\cal O}(K^2)$.

\begin{widetext}

\subsection{Cumulant generating function}\label{}

In order to calculate $\psi_2(\qq,t)$ in Eq.~\eqref{eq:psi-calc-1}, we focus on  $\left\langle \ed^{\id\qq\cdot\YY(t)} S\ind{int}\right\rangle_\ga$.
Taking into account the structure of  $S\ind{int}$ in Eq.~\eqref{eq:Sint-def} and the definitions of $F_\mu$ and $G_{\mu\nu}$ in Eqs.~\eqref{eq:F-def} and \eqref{eq:G-def}, respectively, it emerges that this calculation requires the knowledge of the following quantities, which we determine using Eq.~(35) in Ref.~\cite{Demery2011Perturbative}, i.e.,
\begin{multline}
\left\langle \ed^{\id\qq\cdot\YY(t)} P_\mu(t')\ed^{\id\kk\cdot[\XX(t')-\XX(t'')]} \right\rangle_\ga = -q_\mu[\mcR(t-t')-\mcR(-t')]\\
\qquad\qquad\qquad\times \exp \left(-\frac{1}{2}\left[\sigma_0^2(t)q^2+\sigma_0^2(t'-t'')k^2 \right]-\kk\cdot\qq \mcS(t,t',t'') \right),
\end{multline}
and
\begin{multline}
\left\langle \ed^{\id\qq\cdot\YY(t)} P_\mu(t')P_\nu(t'')\ed^{\id\kk\cdot[\XX(t')-\XX(t'')]} \right\rangle_\ga 
= q_\mu[\mcR(t-t')-\mcR(-t')] \left\{q_\nu[\mcR(t-t'')-\mcR(-t'')]+k_\nu\mcR(t'-t'') \right\}\\
\times \exp \left(-\frac{1}{2}\left[\sigma_0^2(t)q^2+\sigma_0^2(t'-t'')k^2 \right]-\kk\cdot\qq \, \mcS(t,t',t'') \right),
\end{multline}
with $t'>t''$, where introduced
\begin{equation}
\mcS(t,t',t'')=\frac{D_0}{\omega_0}\left[\mcR(t-t')-\mcR(t-t'')-\mcR(|t'|)+\mcR(|t''|) \right].
\end{equation}
Combining these expressions, we obtain for the correction in Eq.~\eqref{eq:psi-calc-1},
\begin{multline}
\label{eq:psiK-fin}
\psi_2(\qq,t) = -\frac{1}{\gamma}\int \frac{\dd\kk}{(2\pi)^d}\frac{\wt K^2}{\wt A}(\kk\cdot\qq)\int\dd t'\dd t''\,\theta(t'-t'')
\ed^{-\wt R\wt A (t'-t'')-\frac{k^2}{2}\sigma_0^2(t'-t'')-\kk\cdot\qq\,\mcS(t,t',t'')}\\
\times \left[\mcR(t-t')-\mcR(-t') \right]\left\{\wt R\wt A+D_0 \left\{\kk\cdot\qq[\mcR(t-t'')-\mcR(-t'')]+k^2\mcR(t'-t'') \right\} \right\}.
\end{multline}
After some simplifications (see App.~\ref{app:simp_cgf} for details), we arrive at 
\begin{equation}\label{eq:cgf2}
\psi_2(q,t) = \frac{2D_0^2q^2}{\omega_0 T}\ed^{-\omega_0 t}\int \frac{\dd\kk}{(2\pi)^d}\frac{k_1^2\wt K^2}{\wt A}\int_0^t\dd s \, \ed^{(\omega_0-\wt R\wt A) s-\frac{k^2}{2}\sigma_0^2(s)}J_0 \left(\frac{4D_0 q}{\omega_0}k_1\ed^{-\omega_0 t/2}\sinh \left(\frac{\omega_0 s}{2} \right) ,\frac{\omega_0(t-s)}{2}\right),
\end{equation}
where $J_0$ is the lower-incomplete form of the modified Bessel function of the second kind:
\begin{equation}\label{eq:lif_bessel}
J_0(z,w)= \int_0^w\!\dd v\, \exp(-z\cosh v).
\end{equation}
Equation~\eqref{eq:cgf2} describes in full generality the statistical effects of 
the probe-field coupling, at the lowest non-trivial perturbative order in that coupling.

\subsection{Moments of the displacement}\label{}

The various moments of the displacement $\YY(t)$ are obtained by expanding $\psi(\qq,t)$ in Eq.~\eqref{eq:psi-calc-1} in powers of $q$, taking into account also Eq.~\eqref{eq:cgf2}. In particular, as the bare term $\psi_0(\qq,t)$ is already quadratic in $q= |\qq|$, non-Gaussian contributions to the kurtosis can be obtained by expanding the correction $\psi_2(q=|\qq|,t)$ in powers of $q$, according to 
\begin{equation}
\label{eq:psiK-exp}
\psi_2(q,t) = -\frac{1}{2}\psitt(t)q^2+\frac{1}{4!}\psitf(t)q^4 + {\cal O}(q^6).
\end{equation}
In order to determine $\psitt$  and $\psitf$ one needs to expand the integral expression of $J_0(z,w)$ in Eq.~\eqref{eq:lif_bessel} in powers of its first argument $z  \propto q k_1$. 
As the resulting terms which are odd powers of $z$ do not contribute to the integral in Eq.~\eqref{eq:cgf2} after integration over $\kk$, we do not need to compute them:
\begin{align}
J_0(z,w) & = \int_0^w\!\dd v \, \left[1+\frac{z^2}{2}(\cosh v)^2 \right]+\mcO(z^4)+\textrm{odd powers of }z\\
& = w + \frac{z^2}{8}[\sinh(2w)+2w]+\mcO(z^4)+\textrm{odd powers of }z.
\end{align}
Substituting this expansion in Eq.~\eqref{eq:cgf2}, we get
\begin{align}
\psitt(t) & = -\frac{2D_0^2}{T}\ed^{-\omega_0 t}\int \frac{\dd\kk}{(2\pi)^d}\frac{k_1^2\wt K^2}{\wt A}\int_0^t\dd s \,  \ed^{(\omega_0-\wt R\wt A) s-\frac{k^2}{2}\sigma_0^2(s)}(t-s),\label{eq:msd_trap0}\\
\psitf(t) & = \frac{96D_0^4 }{T\omega_0^3}\ed^{-2\omega_0 t}\int \frac{\dd\kk}{(2\pi)^d}\frac{k_1^4\wt K^2}{\wt A}\int_0^t\dd s \, \ed^{(\omega_0-\wt R\wt A) s-\frac{k^2}{2}\sigma_0^2(s)}\sinh^2 \left(\frac{\omega_0 s}{2} \right)\left[\sinh(\omega_0(t-s))+\omega_0(t-s)\right], \label{eq:psi2-fin0}
\end{align}
Under the assumption of an isotropic system, the factors $k^2_1$ in Eq.~\eqref{eq:msd_trap0} and $k^4_1$ in Eq.~\eqref{eq:psi2-fin0} can be replaced by $k^2/d$ and $3k^4/[d(d+2)]$, respectively.
The latter follows from the fact that a rotationally-invariant integral of the product $k_\mu k_\nu k_\sigma k_\tau$ has to be proportional to $\delta_{\mu\nu}\delta_{\sigma\tau}+\delta_{\mu\sigma}\delta_{\nu\tau}+\delta_{\mu\tau}\delta_{\nu\sigma}$.
The remaining integral over the angle of $\kk$ renders the $d$-dimensional solid angle $\Omega_d = 2\pi^{d/2}/\Gamma(d/2)$, leading to:
\begin{align}
\psitt(t) & = -\chi_2 \frac{D_0^2}{T}\ed^{-\omega_0 t}\int_0^\infty\dd k \frac{k^{d+1}\wt K^2}{\wt A}\int_0^t\dd s \,  \ed^{(\omega_0-\wt R\wt A) s-\frac{k^2}{2}\sigma_0^2(s)}(t-s),\label{eq:msd_trap}\\
\psitf(t) & = \chi_4\frac{D_0^4 }{T\omega_0^3}\ed^{-2\omega_0 t}\int_0^\infty \dd k \frac{k^{d+3}\wt K^2}{\wt A}\int_0^t\dd s \, \ed^{(\omega_0-\wt R\wt A) s-\frac{k^2}{2}\sigma_0^2(s)}\sinh^2 \left(\frac{\omega_0 s}{2} \right)\left[\sinh(\omega_0(t-s))+\omega_0(t-s)\right], \label{eq:psi2-fin}
\end{align}
where we have introduced the numerical factors
\begin{align}
\chi_2& = \frac{2\Omega_d}{d(2\pi)^d} = \frac{4}{d(4\pi)^{d/2}\Gamma(d/2)},\\
\chi_4& = \frac{288\Omega_d}{d(d+2)(2\pi)^d}= \frac{576}{d(d+2)(4\pi)^{d/2}\Gamma(d/2)}.
\end{align}
Using these expressions into Eqs.~\eqref{eq:psiK-exp} and \eqref{eq:psi-calc-1}, the resulting $\psi(\qq,t)$ can be expanded as in Eq.~\eqref{eq:psi-dec}, finding
\begin{align}
\psi^{(2)}(t) &= \sigma_0^2(t) + \psitt(t) + {\cal O}(K^4),\\
\psi^{(4)}(t) &= \psitf(t) + {\cal O}(K^4). \label{eq:psi-dec-pow}
\end{align}
Accordingly, the excess kurtosis in Eq.~\eqref{eq:def-gamma} is given, neglecting ${\cal O}(K^4)$, by
\begin{multline}\label{eq:kurt_trap}
\gamma(t) = 
\frac{\psi^{(4)}(t)}{\left[ \psi^{(2)}(t)\right]^2} = \frac{\psitf(t)}{\left[\sigma_0^2(t)\right]^2} 
\\=\frac{\chi_4}{4}\frac{D_0^2 }{T\omega_0}\frac{1}{ \left(\ed^{\omega_0 t}-1 \right)^2}\int_0^\infty\dd k \frac{k^{d+3}\wt K^2}{\wt A}\int_0^t\dd s \, \ed^{(\omega_0-\wt R\wt A) s-\frac{k^2}{2}\sigma_0^2(s)}\sinh^2 \left(\frac{\omega_0 s}{2} \right)\left[\sinh(\omega_0(t-s))+\omega_0(t-s)\right],
\end{multline}
where we have used Eqs.~\eqref{eq:def-sig0} and \eqref{eq:psi2-fin}.
Equation \eqref{eq:msd_trap} for the mean-square displacement and Eq.~\eqref{eq:kurt_trap} for the excess kurtosis are the main predictions of this work.

\subsection{Free probe}\label{}

The case of a freely diffusing probe can be obtained by setting the stiffness $\kappa$ of the trap to zero and thus
by considering $\omega_0\to 0$ (see the definition of $\omega_0$ after Eq.~\eqref{eq:dyn_probe}) and $\sigma_0^2(t) \to 2D_0t$ (see Eq.~\eqref{eq:def-sig0}).
The second moment in Eq.~\eqref{eq:msd_trap} becomes, in this case,
\begin{equation}
\psitt(t) = -\frac{\chi_2 D_0^2}{T}\int_0^\infty\dd k\frac{k^{d+1}\wt K^2}{\wt A\alpha^2}\left(\alpha t-1+\ed^{-\alpha t} \right),
\label{eq:psiK2-free}
\end{equation}
where we have introduced the relaxation rate
\begin{equation}
\alpha(k)=\wt R(k)\wt A(k)+D_0k^2.
\label{eq:def-alpha}
\end{equation}
The ``effective diffusion coefficient'' is naturally defined as
\begin{equation}
D(t)= \frac{\langle Y_1^2(t)\rangle}{2t}= \frac{\psi_2(t)}{2t},
\label{eq:Deff}
\end{equation}
where we used Eq.~\eqref{eq:def-psi2}.
Taking into account Eqs.~\eqref{eq:psi-dec-pow} and \eqref{eq:psiK2-free}, from the latter expression above, one finds at order $K^2$, that
\begin{equation}\label{eq:effective_diff_coeff}
\frac{D(t)}{D_0}-1 = \frac{\psitt(t)}{2D_0 t} 
= -\frac{\chi_2 D_0}{2T}\int_0^\infty\dd k\frac{k^{d+1}\wt K^2}{\wt A\alpha^2}\left(\alpha-\frac{1-\ed^{-\alpha t} }{t}\right).
\end{equation}
For the fourth moment, instead, from Eq.~\eqref{eq:psi2-fin} we have
\begin{align}
\psitf(t) &= \frac{\chi_4 D_0^4 }{2T}\int_0^\infty \dd k\frac{k^{d+3}\wt K^2}{\wt A}\int_0^t\dd s\,  \ed^{-\alpha s}s^2(t-s)\\
& = \frac{\chi_4 D_0^4 }{2T}\int_0^\infty\dd k \frac{k^{d+3}\wt K^2}{\wt A\alpha^4}\left[2\alpha t-6+(\alpha^2t^2+4\alpha t+6)\ed^{-\alpha t}\right].
\end{align}
Accordingly, the excess kurtosis at order $K^2$ is given by
\begin{equation}\label{eq:kurtosis_free}
\gamma(t) = \frac{\psitf(t)}{(2D_0t)^2} = \frac{\chi_4 D_0^2 }{8Tt^2}\int_0^\infty \dd k \frac{k^{d+3}\wt K^2}{\wt A\alpha^4}\left[2\alpha t-6+(\alpha^2t^2+4\alpha t+6)\ed^{-\alpha t}\right].
\end{equation}
In Sec.~\ref{sec:asymptotic} we shall discuss the qualitative features of the dependence of $D(t)$ and $\gamma(t)$ above on time $t$ for some specific choices of the operators $A$, $K$, and $R$, which define the model in full generality. 
In the next section, instead, we focus on the comparison between our theoretical predictions and the numerical simulations  of the stochastic dynamics of the system.

\end{widetext}

\section{Numerical simulations}
\label{sec:simulations}

\subsection{Model}\label{}

In order to compare the perturbative predictions obtained in the previous sections with the (non-perturbative) results of numerical simulations, we focus on the specific model with $\wt A(k)=k^2$, $\wt R(k)=1$, $\wt K(k)=h=0.5$, $D_0=1$, and $T=1$, in spatial dimension $d=1$. 
The stochastic dynamics of the field (Eq.~\eqref{eq:dyn_field}) is simulated in Fourier space~\cite{Kraichnan1976Diffusion, Demery2011Perturbative} with a finite number of modes with wavevectors $k = 1,\dots, N$; here we take $N=10$.
The stochastic dynamics of the probe is simulated in real space; the force due to the field --- given by second term on the r.h.s.~of Eq.~\eqref{eq:dyn_probe} --- is evaluated at each time step from the Fourier coefficients of the field.
This method avoids using a lattice model, which would require a discretization of the field and an interpolation step to compute its effect on the (off-lattice) particle.
Note that, taking into account the smallest wavevector and counting the numbers of degrees of freedom, the present numerical approach corresponds to a discretization on a lattice with spacing $\pi/L$, extension $L=2\pi$, and periodic boundary conditions.
The stochastic differential equations for the position of the particle and for the coefficients of the Fourier modes of the field are integrated with an $o(\delta t^2)$ scheme~\cite{Drummond1986Numerical}.

The model chosen here would correspond, in an infinitely extended system, to a \emph{critical} Gaussian model for the field, with non-conserved dynamics and a localized $\delta$-interaction with the probe.
However, the finite size of the system, encoded in the smallest wavevector $k=1$, makes the field practically non-critical with correlation length $\xi\simeq 2\pi$.

Our perturbative, analytical predictions for the second- and fourth-order moments of the displacement and therefore for the excess kurtosis can be adapted to the case in which one considers a finite number of modes in the numerical simulations. 
In particular, for $d=1$ the integrals over the modes in Eqs.~\eqref{eq:msd_trap}, \eqref{eq:psi2-fin}, and  \eqref{eq:kurt_trap} are replaced by discrete sums as
\begin{equation}
\int_0^\infty\dd k \longrightarrow \sum_{k=1}^N.
\end{equation}

\subsection{Numerical results}\label{}

\begin{figure}
\begin{center}
\includegraphics[scale=1]{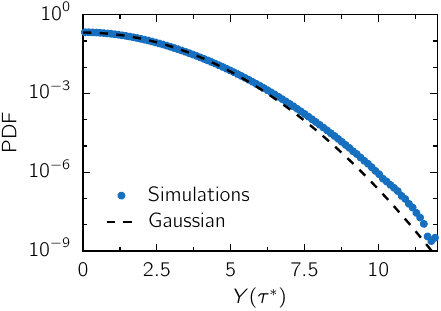}
\end{center}
\caption{Probability distribution function (PDF) of the displacement $Y(\tau^*=2.21)$ at the time $\tau^*$ at which the excess kurtosis attains its maximum.
Symbols correspond to the result of numerical simulations in $d=1$ with $N=10$ modes for $\wt A(k)=k^2$, $\wt R(k)=1$, $\wt K(k)=1$, $D_0=1$, $T=1$ and $h=1$ without trap, while the dashed line indicates a Gaussian distribution with the same standard deviation as the numerical data. The discrepancy between symbols and the dashed line highlights the non-Gaussian nature of the fluctuations of the particle position also beyond perturbation theory.}
\label{fig:pdf}
\end{figure}

Figure~\ref{fig:pdf}  shows,  in the absence of the trap, the probability distribution function (PDF) of the displacement $Y(t=\tau^*)$ in a time interval  $\tau^*=2.21$, where $\tau^*$ is the value of the time $t$ at which the excess kurtosis $\gamma(t)$ of the displacement is maximal (see below).
The PDF obtained from our numerical simulations is compared with a Gaussian distribution with the same variance, showing that the former is Gaussian for small values of the displacement while its tails are slightly heavier than those of a Gaussian distribution.
In order to characterize this PDF in more detail, we now analyze its lowest cumulants: the mean squared displacement and the excess kurtosis.

\begin{figure}
\includegraphics[scale=1]{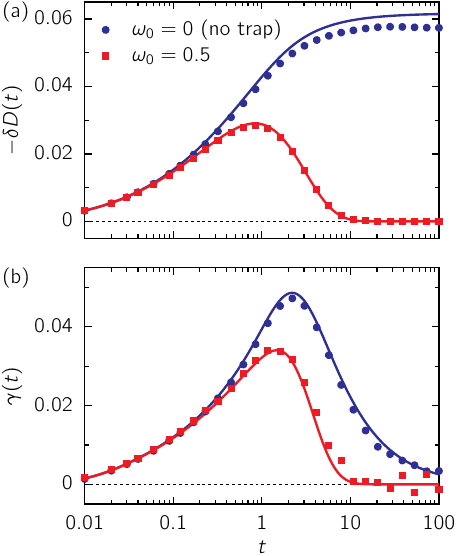}
\caption{
Comparison between the analytic perturbative predictions at order $K^2$ (solid lines) and the numerical simulations (symbols) in $d=1$ with $N=10$ modes, for $\wt A(k)=k^2$, $\wt R(k)=1$, $\wt K(k)=1$, $D_0=1$, $T=1$, and $h=0.5$ in the absence (blue circles) or in the presence  (red squares) of a trap with relaxation rate $\omega_0=0.5$. 
(a) Correction $\delta D(t)$ to the diffusion coefficient $D(t)$ (see Eqs.~\eqref{eq:Deff} and \eqref{eq:DeltaD}) as a function of time $t$. The analytical predictions are given by Eq.~\eqref{eq:effective_diff_coeff} in the absence of the trap and by Eq.~\eqref{eq:msd_trap} in its presence. 
(b) Excess kurtosis $\gamma(t)$ as a function of time $t$. The corresponding analytical predictions are given by Eq.~\eqref{eq:kurtosis_free} in the absence of the trap and by Eq.~\eqref{eq:kurt_trap} in its presence.
}
\label{fig:msd_kurt}
\end{figure}

%
We first consider the correction to the time-dependent diffusion coefficient $D(t)$ which, in the presence of a potential,  can still be formally defined as in Eq.~\eqref{eq:Deff} in terms of the second moment $\langle Y^2(t)\rangle $ of the displacement $Y(t)$.
According to Eq.~\eqref{eq:psi-dec-pow}, in the absence of the interaction with the field, $D(t)$ takes the bare value $D_0(t) = \sigma_0^2(t)/(2t)$ with $\sigma_0^2$ given in Eq.~\eqref{eq:def-sig0}. 
Accordingly, their difference 
\begin{equation}
\delta D(t) = D(t)- D_0(t) =  D(t) - \frac{\sigma_0^2(t)}{2t},
\label{eq:DeltaD}
\end{equation}
can be calculated via numerical simulations and can be compared with its theoretical perturbative estimate at ${\cal O}(K^2)$ given by $\psi_2^{(2)}(t)/(2t)$ with $\psi_2^{(2)}$ provided in Eq.~\eqref{eq:msd_trap}.
The results obtained from numerical simulations in the absence or in the presence of the trap with relaxation rate $\omega_0=0.5$ are shown in Fig.~\ref{fig:msd_kurt}(a), with blue circles and red squares, respectively. 
The solid lines with the corresponding colours are the perturbative analytical predictions. 
The agreement between numerical data and theoretical predictions is very good, except that for a free probe at large times, where the theoretical prediction overestimates the actual correction; this small difference might be due to the fact that the actual value of $\wt K(k)=h$ is large enough to require the inclusion of corrections of order $K^4$ and higher, which are currently neglected in our ${\cal O}(K^2)$ theoretical prediction. 

The excess kurtosis $\gamma(t)$ is defined in Eq.~\eqref{eq:def-gamma} in terms of the cumulants of the displacement $Y(t)$.
In Fig.~\ref{fig:msd_kurt}(b)  we show the results of our numerical simulations in the absence or in the presence of trap with relaxation rate $\omega_0=0.5$, with blue circles and red squares, respectively. 
As in Fig.~\ref{fig:msd_kurt}(a), the solid lines with the corresponding colours are the perturbative analytical predictions for $\gamma(t)$ in Eqs.~\eqref{eq:kurt_trap} and \eqref{eq:kurtosis_free}.  The agreement between the numerical data and the predictions turns out to be very good both in the presence and in the absence of the trap. 

The numerical evidence presented here shows that our predictions for the mean squared displacement and for the kurtosis 
are in good agreement with the results of the simulations. However, we cannot make a prediction for the full probability distribution of the displacement, because this would require the knowledge of all the cumulants.

\section{Asymptotic behaviors}
\label{sec:asymptotic}

In this section we analyze the asymptotic behaviors at short and long times of the correction to the mean-square displacement and to the kurtosis.
For concreteness, we consider a Gaussian field with the energy given by $A$ in Eq.~\eqref{eq:A-Gaux}, corresponding to 
\begin{equation}
\wt A(\kk)=k^2+m^2,
\label{eq:ex-A} 
\end{equation}
where $m$ is the inverse of the spatial correlation length $\xi$ of the fluctuations of the field, with $\widetilde R(k)=rk^\rho$, where $\rho=0$ corresponds to a non-conserved dynamics (model A), and $\rho=2$ to a conserved dynamics (model B)~\cite{Hohenberg1977}.
The parameter $r$ is a diffusion coefficient for $\rho=0$ and has the dimension $\mathrm{length}^4/\mathrm{time}$ for $\rho=2$.
The probe-field interaction is assumed to be given by $\wt K(k)=h\exp(-a^2k^2/2)$, 
where $a$ plays the role of the size of the probe.

Depending on the spatial dimensionality $d$ and on the correlation length $\xi$, the Fourier integrals involved in Eqs.~\eqref{eq:msd_trap} and \eqref{eq:kurt_trap} 
may in principle develop 
infrared and/or ultraviolet divergences. 
The ultraviolet divergences are actually avoided by considering a probe with a finite linear extension $a$, which eventually controls the value of these integrals. The infrared divergences, instead, are more interesting for the present study, as they may affect the long-time behavior of the quantities considered here.

\subsection{Short-time behavior} \label{sec:short}

At short times, the correction $\psitt(t)$ in Eq.~\eqref{eq:msd_trap}  to the MSD and the excess kurtosis in Eq.~\eqref{eq:kurt_trap} scale, respectively,  as
\begin{align}
\psitt(t)
&\underset{t\to 0}{\sim} -\frac{\chi_2 D_0^2}{2T}t^2\int_0^\infty \dd k\frac{k^{d+1}\wt K^2}{\wt A} ,\\
\gamma(t) &\underset{t\to 0}{\sim} \frac{\chi_4 D_0^2 }{96T}t^2\int_0^\infty \dd k\frac{k^{d+3}\wt K^2}{\wt A} . \label{eq:kurt_short} 
\end{align}
It is easy to see that the resulting integrands have no divergence for $k\to 0$
and therefore one concludes that both $\psitt(t)$ and $\gamma(t)$ initially grow as $t^2$ upon increasing $t$.
Note that these integrands are actually independent of $\wt R(k)$, i.e., of the specific dynamics of the field, while they are determined by its \emph{static} properties and by the coupling between the field and the particle, encoded, respectively, in $\wt A$ and $\wt K$.
Note also that this short-time behavior does not depend on the presence of the trap.

\begin{widetext}

\subsection{Long-time behavior}
\label{sec:long-time}

The long-time behaviour of the effective diffusion coefficient and of the excess kurtosis turn out to depend on the possible presence of the trap and therefore we discuss separately these two cases  below.

\subsubsection{In the absence of the trap}

In the absence of the trap, the correction $\delta D(t)$ to the diffusion coefficient  reaches its asymptotic value $\delta D_\infty$ at long times, with
\begin{equation}
\label{eq:Dinfty}
\delta D_\infty = \lim_{t\to\infty} \delta D(t) = -\frac{\chi_2 D_0^2}{2T}\int_0^\infty\dd k\frac{k^{d+1}\wt K^2}{\wt A\alpha},
\end{equation}
where $\alpha$ is given in Eq.~\eqref{eq:def-alpha}.
If the correlation length $m^{-1}$ in Eq.~\eqref{eq:ex-A} is finite, this integral converges. If the correlation length is infinite, i.e., for $m=0$, instead, the integral converges only for $d>2$.
Note that $\delta D_\infty$ provides a \emph{negative} correction to the bare diffusion coefficient $D_0$ and that $|\delta D_\infty|$ increases upon increasing the strength of $K$. As the long-time diffusion coefficient $D_\infty= D_0+\delta D_\infty$ has to be positive, this requires that $0<-\delta D_\infty<D_0$, i.e., that the perturbative expansion up to ${\cal O}(K^2)$ developed here has to fail for a sufficiently strong particle-field interaction strength $K$. 
In order to understand the behavior of $\delta D$ when this integral, instead, diverges  (i.e., for $d\le 2$ and $m=0$)
we change the integration variable to $q=\sqrt{t}k$ in Eq.~\eqref{eq:effective_diff_coeff}, leading to
\begin{equation}
\delta D(t)=D(t)-D_0=-\frac{\chi_2 h^2D_0^2}{2T}t^{1-d/2}\int_0^\infty\dd q \frac{q^{d-5} \ed^{-a^2 q^2/(2t)}}{\left(D_0+rq^\rho/t^{\rho/2} \right)^2}\left[q^2 \left(D_0+\frac{rq^\rho}{t^{\rho/2}} \right)-1+\ed^{q^2 \left(D_0+\frac{rq^\rho}{t^{\rho/2}} \right)} \right].
\label{eq:asymp-free-D}
\end{equation}
In spatial dimensionality $d=1$, introducing $D_0'=D_0+r$ 
for model A ($\rho=0$) and $D_0'=D_0$ for model B ($\rho=2$), the long-time behavior turns out to be given by
\begin{equation}
\delta D(t) \underset{t\to\infty}{\sim}  -\frac{h^2D_0^2}{\pi T\sqrt{D_0'}}\sqrt{t}\int_0^\infty\dd p \frac{p^2-1+\ed^{-p^2}}{p^4}
=-\frac{2}{3\sqrt{\pi}}\frac{h^2D_0^2}{T\sqrt{D_0'}}\sqrt{t} .
\end{equation}
For $d=2$, instead, taking into account the finite size $a$ of the particle, Eq.~\eqref{eq:asymp-free-D} renders
\begin{equation}
\delta D(t) \underset{t\to\infty}{\sim} -\frac{1}{8\pi}\frac{h^2D_0^2}{T D_0'}\ln (t) ,
\end{equation}
where $a$ enters the inconsequential multiplicative factor which makes the argument of the logarithm dimensionless.
As already discussed above, the diffusion coefficient $D(t)$ is positive by its very definition (see Eq.~\eqref{eq:Deff}) and therefore the correction $\delta D(t)$ to the diffusion coefficient, being generically \emph{negative}, cannot grow forever. Accordingly, at some time $t^*$ such that $-\delta D(t^*) \simeq D_0$ determined from the previous equations, the perturbative calculation has to break down.  
Note that  the increase of $\delta D(t)$ upon increasing time for $d\leq 2$ may indicate the possible occurrence of anomalous diffusion.
The different behaviors of the correction $\psitt(t)$ to the mean squared displacement depending on the model, spatial dimensionality, and value of $m$ are summarized in Table~\ref{tab:long_psi2}.

The excess kurtosis in the absence of the trap is given by Eq.~\eqref{eq:kurtosis_free} and at long times it behaves as
\begin{equation}
\gamma(t)  \underset{t\to\infty}{\sim}  \frac{\chi_4 D_0^2 }{4Tt}\int_0^\infty \dd k\frac{k^{d+3}\wt K^2}{\wt A\alpha^3},
\end{equation}
i.e., $\gamma(t)\sim t^{-1}$, provided that the integral converges, which is the case if $m>0$ and $d>3\rho-4$ or if $m=0$ and $d>4$. 
The behavior of $\gamma(t)$ is different if the integral diverges, i.e., for $m=0$ and $d\le 4$ in both model A and model B or for model B with $d\le 2$ and $m\neq 0$.
In both cases, the relaxation rate $\alpha$ given in Eq.~\eqref{eq:def-alpha} scales as $\alpha(k)\simeq D_0' k^2$, for small values of $k$, with $D_0'=D_0+r$ for model A and $m=0$, $D_0'=D_0$ for model B and $m=0$, and $D_0'=D_0+rm^2$ for model B and $m>0$.

For $m=0$, changing the integration variable to $q=\sqrt{D_0' t}k$ in Eq.~\eqref{eq:kurtosis_free} yields, for $d<4$,
\begin{equation}
\gamma(t)\underset{t\to\infty}{\sim} \frac{\chi_4 D_0^2 h^2}{8T{D_0'}^{1+d/2}}t^{1-d/2}\int_0^\infty \frac{\dd q}{q^{7-d}}\left[2q^2-6+(q^4+4q^2+6)\ed^{-q^2} \right].
\label{eq:kurt-app-1}
\end{equation}
For $d=4$, we get
\begin{equation}
\gamma(t) \underset{t\to\infty}{\sim} \frac{3}{16\pi^2} \frac{D_0^2 h^2}{T{D_0'}^3}\frac{\ln (t)}{t}.
\end{equation}
Using the same manipulations as above for the case $m>0$ and model B (i.e., with $\rho=2$), we find, for $d=1$,
\begin{equation}
\gamma(t) \underset{t\to\infty}{\sim}   \frac{6}{\sqrt{\pi}}\frac{D_0^2 h^2}{Tm^2 {D_0'}^{5/2}}t^{-1/2},
\end{equation}
while, for $d=2$,
\begin{equation}
\gamma(t) \underset{t\to\infty}{\sim}  \frac{9}{4\pi} \frac{D_0^2 h^2}{Tm^2{D_0'}^3} \frac{\ln (t)}{t}.
\end{equation}
The different behaviors of the excess kurtosis depending on the model, spatial dimension and on the value of $m$ are summarized in Table~\ref{tab:long_kurt}.

\begin{table}
\begin{center}
\begin{tabular}{|l|c|c|c|c|c|c|}
\cline{2-7}
\multicolumn{1}{c|}{}&\multicolumn{4}{c|}{Model A} & \multicolumn{2}{c|}{Model B}\\\cline{2-7}
\multicolumn{1}{c|}{}&\multicolumn{2}{c|}{$m>0$} & \multicolumn{2}{c|}{$m=0$} & $m>0$ & $m=0$ \\\hline
\multirow{3}{*}{no trap}& \multirow{3}{*}{any $d$}& \multirow{3}{*}{$t$}& $d=1$ & $t^{3/2}$  & \multicolumn{2}{c|}{\multirow{3}{*}{same as model A}} \\\cline{4-5}
                        &                         &                     & $d=2$ & $t\ln (t)$ & \multicolumn{2}{c|}{}   \\\cline{4-5}
                        &                         &                     & $d>2$ & $t$        & \multicolumn{2}{c|}{}   \\\hline
with trap               &                         & $\ed^{-rm^2t}$      &       & $t^{-d/2}$ &               $t^{-(d+2)/2}$       & $t^{-d/4}$ \\\hline
\end{tabular}
\end{center}
\caption{Summary of the long-time behavior of the correction $\psitt(t)=2t\delta D(t)$ to the mean squared displacement, as a function of time $t$, discussed in Sec.~\ref{sec:long-time}. 
}
\label{tab:long_psi2}
\end{table}

\begin{table}
\begin{center}
\newcommand{\minitab}[2][l]{\begin{tabular}{#1}#2\end{tabular}}
\begin{tabular}{|l|c|c|c|c|c|c|c|}
\cline{2-8}
\multicolumn{1}{c|}{}&\multicolumn{4}{c|}{Model A} & \multicolumn{3}{c|}{Model B}\\\cline{2-8}
\multicolumn{1}{c|}{}&\multicolumn{2}{c|}{$m>0$} & \multicolumn{2}{c|}{$m=0$} & \multicolumn{2}{c|}{$m>0$} & $m=0$ \\\hline
\multirow{3}{*}{no trap} & \multirow{3}{*}{any $d$} & \multirow{3}{*}{$t^{-1}$} & $d<4$ & $t^{-(d-2)/2}$ & $d=1$ & $t^{-1/2}$ &  \multirow{3}{*}{\minitab[c]{same as\\ model A}}   \\\cline{4-7}
 &  &  & $d=4$ & $\ln(t)/t$ & $d=2$ & $\ln(t)/t$ &  \\\cline{4-7}
  &  &  & $d>4$ & $t^{-1}$ & $d>2$ & $t^{-1}$ &   \\\hline
with trap &  & $\ed^{-rm^2t}$ &  & $t^{-(d+2)/2}$ &  & $t^{-(d+4)/2}$ &  $t^{-(d+2)/4}$ \\\hline
\end{tabular}
\end{center}
\caption{Summary of the long-time behavior of the excess kurtosis $\gamma(t)$ as a function of time $t$, discussed in Sec.~\ref{sec:long-time}. 
}
\label{tab:long_kurt}
\end{table}

\subsubsection{In the presence of the trap}\label{}

In the presence of the trap,
the correction $\psitt(t)$ to the mean-square displacement is dominated by $s\simeq t$ in the integrand of Eq.~\eqref{eq:msd_trap}, leading to
\begin{align}
\psitt(t) &= -\frac{\chi_2 D_0^2}{T}\int_0^\infty \dd k \frac{k^{d+1}\wt K^2}{\wt A}\int_0^t\dd s\, \ed^{-\omega_0 (t-s)} (t-s)\ed^{-\wt R\wt A s- k^2 \sigma_0^2(s)/2}\nonumber\\
& \simeq -\frac{\chi_2 D_0^2}{T\omega_0^2} \int_0^\infty \dd k\frac{k^{d+1}\wt K^2}{\wt A}\ed^{-\wt R\wt A t- k^2 D_0/\omega_0},
\label{eq:psiK2-2}
\end{align}
where we have also used that $\sigma_0^2(t)\to 2D_0/\omega_0$ as $t\to\infty$, see Eq.~\eqref{eq:def-sig0}.
Depending on the values of $m$ (positive or zero) and $\rho$, suitable changes of variables allow one to determine the algebraic long-time behavior of Eq.~\eqref{eq:psiK2-2}.

In particular, for $\rho=0$ (model A),  the decay upon increasing $t$ is exponential with rate $rm^2$ if $m>0$. If $m=0$, instead,
we define $q=\sqrt{rt}k$, finding
\begin{equation}
\psitt(t)\underset{t\to\infty}{\sim} -\frac{\chi_2 D_0^2 h^2}{T \omega_0^2r^{d/2}}\frac{1}{t^{d/2}}\int_0^\infty\dd q\, q^{d-1}\ed^{-q^2}
=-\frac{\chi_2\Gamma(d/2)}{2} \frac{D_0^2 h^2}{T \omega_0^2r^{d/2}}\frac{1}{t^{d/2}}.
\end{equation}
For $\rho=2$ (model B) and $m>0$, using $q=m\sqrt{rt}k$ leads to
\begin{equation}
\psitt(t)\underset{t\to\infty}{\sim} -\frac{\chi_2 D_0^2 h^2}{T \omega_0^2 m^{4+d}r^{(d+2)/2}}\frac{1}{t^{(d+2)/2}}\int_0^\infty\dd q\, q^{d+1}\ed^{-q^2}
= -\frac{\chi_2 \Gamma(1+d/2)}{2}\frac{D_0^2 h^2}{T \omega_0^2 m^{4+d}r^{(d+2)/2}}\frac{1}{t^{(d+2)/2}}.
\end{equation}
If, instead, for the same value of $\rho=2$ one consider the critical case $m=0$, we define $q=(rt)^{1/4}k$ and get
\begin{equation}
\psitt(t)\underset{t\to\infty}{\sim} -\frac{\chi_2 D_0^2 h^2}{T \omega_0^2 r^{d/4}}\frac{1}{t^{d/4}}\int_0^\infty\dd q\, q^{d-1}\ed^{-q^2}
=-\frac{\chi_2\Gamma(d/2)}{2}\frac{\chi_2 D_0^2 h^2}{T \omega_0^2 r^{d/4}}\frac{1}{t^{d/4}}.
\end{equation}
Note that the correlations for model B ($\rho=2$) have already been obtained in Ref.~\cite{Basu2022Dynamics}.

Concerning the excess kurtosis in Eq.~\eqref{eq:kurt_trap}, the same manipulations as above yield
\begin{align}
\gamma(t)&\simeq \frac{\chi_4 D_0^2}{16\omega_0 T}\int_0^\infty\dd k  \frac{k^{d+3}\wt K^2}{\wt A}\int_0^t \dd s\,\ed^{-2\omega_0(t-s)} \left[\sinh(\omega_0[t-s])+\omega_0(t-s) \right]\ed^{-\wt R\wt A s- k^2\sigma_0^2(s)/2}, \nonumber\\
&\simeq \frac{7\chi_4 D_0^2}{192\omega_0^2 T}\int_0^\infty \dd k\frac{k^{d+3}\wt K^2}{\wt A}\ed^{-k^2 D_0/\omega_0-\wt R\wt A t}.
\end{align}
From this expression, we obtain that the excess kurtosis $\gamma(t)$ for $\rho=0$ (model A) decays exponentially with rate $rm^2$ upon increasing $t$ if $m>0$. 
If, instead, $m=0$, it decays algebraically as 
\begin{equation}
\gamma(t)\underset{t\to\infty}{\sim} \frac{7\chi_4  D_0^2h^2}{192\omega_0^2 Tr^{(d+2)/2}} \frac{1}{t^{(d+2)/2}}\int_0^\infty \dd q\, q^{d+1}\ed^{-q^2}
=\frac{7\chi_4 \Gamma(1 + d/2)}{384}\frac{D_0^2h^2}{\omega_0^2 Tr^{(d+2)/2}} \frac{1}{t^{(d+2)/2}}.
\end{equation}
For $\rho=2$ (model B), one gets a generic algebraic decay in the non-critical case $m>0$, i.e., 
\begin{equation}
\gamma(t)\underset{t\to\infty}{\sim}\frac{7\chi_4 D_0^2h^2}{192\omega_0^2 T m^{6+d}r^{(d+4)/2}} \frac{1}{t^{(d+4)/2}} \int_0^\infty \dd q\, q^{d+3}\ed^{-q^2}
=\frac{7\chi_4\Gamma(2 + d/2)}{384}\frac{ D_0^2h^2}{\omega_0^2 T m^{6+d}r^{(d+4)/2}} \frac{1}{t^{(d+4)/2}},
\end{equation}
which changes, for the critical case $m=0$, into
\begin{equation}
\gamma(t)\underset{t\to\infty}{\sim} \frac{7\chi_4 D_0^2h^2}{192\omega_0^2 Tr^{(d+2)/4}} \frac{1}{t^{(d+2)/4}} \int_0^\infty \dd q\, q^{d+1}\ed^{-q^2}
=\frac{7\chi_4\Gamma(1+ d/2)}{384}\frac{D_0^2h^2}{\omega_0^2 Tr^{(d+2)/4}} \frac{1}{t^{(d+2)/4}}.
\end{equation}
The various behaviors of the excess kurtosis $\gamma(t)$ at long times $t$ in the presence of a trap are summarized in Table~\ref{tab:long_kurt}.

In all cases discussed above and summarized in the table, the decay of $\gamma(t)$ upon increasing $t$ in the presence of a trap is faster than without it.
In the absence of a trap, $\gamma(t)$ for critical model A in $d<2$ actually grows in time (see Table~\ref{tab:long_kurt} and Fig.~\ref{fig:numkurt}).
This growth actually occurs within the range of validity of our perturbation theory.
Note that  the prefactors of the asymptotic expressions of $\gamma(t)$ in the presence of a trap depend on $D_0$ and $\omega_0$ only via their ratio $D_0/\omega_0=T/\kappa$.
This means that the mobility $\gamma^{-1}$ of the probe, which enters both $D_0$ and $\omega_0$,  does not play any role.
As for the correlation functions studied in Ref.~\cite{Basu2022Dynamics}, the long-time algebraic decay in the presence of a trap reflects the dynamics of the field, and not an interplay between the dynamics of the probe and that of the field itself.
Conversely, the algebraic decay of the excess kurtosis without the trap involves the interplay between the dynamics of the probe and that of the field.

\end{widetext}

\subsection{Comparison with the numerical integration}\label{}

\begin{figure*}
\begin{center}
\includegraphics[scale=1]{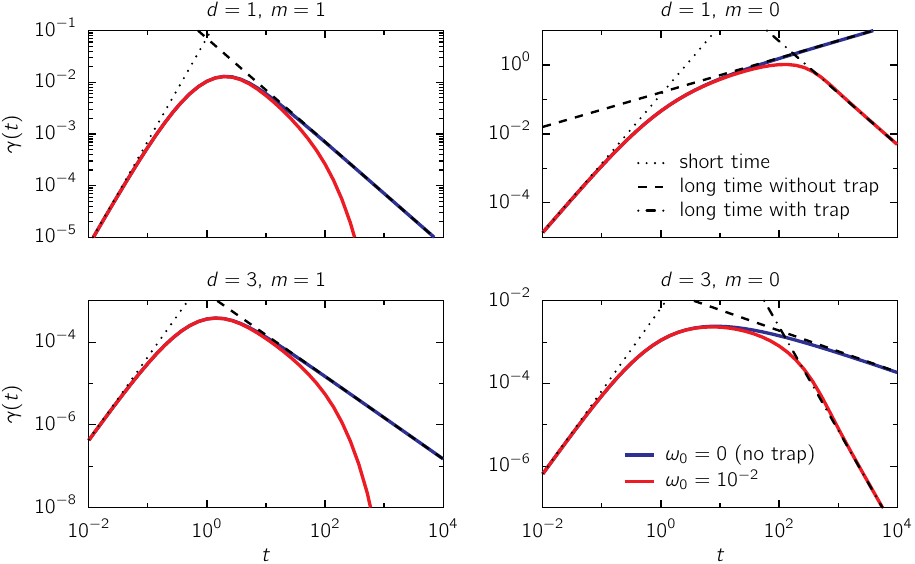}
\end{center}
\caption{
Perturbative prediction of the dependence of the excess kurtosis $\gamma(t)$ on time $t$ in the presence (red, with $\omega_0=10^{-2}$) or in the absence (blue) of a trap, obtained from the numerical integration of Eqs.~\eqref{eq:kurt_trap} and \eqref{eq:kurtosis_free}, respectively.  In particular, we consider the case of  the field with model A dynamics in spatial dimensions $d=1$ (upper panels) and 3 (lower panels), either non-critical ($m=1$, left panels) or critical ($m=0$, right panels).
The dotted lines show the short-time behavior of $\gamma(t)$ (see Eq.~\eqref{eq:kurt_short}), the dash-dotted line corresponds to  the possibly algebraic long-time behavior in the presence of a trap, while the dashed line corresponds to the same limit but in the absence of a trap, see also Tab.~\ref{tab:long_kurt}.
}
\label{fig:numkurt}
\end{figure*}

Figure~\ref{fig:numkurt} shows the perturbative analytical predictions for the evolution of the excess kurtosis $\gamma(t)$ with  (red lines) and without  (blue lines) trap obtained by the numerical integration of Eqs.~\eqref{eq:kurt_trap} and \eqref{eq:kurtosis_free}, respectively, for model A. The various parameters of  the model used for the calculation are $h =1$, $D_0=1$, $a=1$, $T=1$, $d = 1$ (upper panels) or 3 (lower panels) and $m=0$ (left panels) or 1 (right panels). These predictions are compared with the corresponding short- and long-time behaviors derived in Secs.~\ref{sec:short} and \ref{sec:long-time} (see also Tab.~\ref{tab:long_kurt}), which turn out to provide excellent approximations.

\section{Conclusion and perspectives}
\label{sec:conclusion}

In this work, we have  computed perturbatively the second- and fourth-order moments of the displacement of a probe particle coupled to a Gaussian field and held in a harmonic trap.
Our analytical predictions are in good agreement with the results of numerical simulations in which the field is represented by a finite number of Fourier modes.
We have shown that the excess kurtosis $\gamma(t)$ --- which characterizes the effective non-linearity of the probe dynamics --- is positive, vanishes at short times, increases upon increasing time, reaches a maximum and then it generically decreases, vanishing asymptotically at long times. 
This means that the displacement of the probe is Gaussian at short and long times, while its distribution deviates from a Gaussian at intermediate times, similarly to what is observed 
for a colloid in a bacterial bath~\cite{Lagarde2020}. In the absence of the trap and in a critical field in spatial dimension $d< 2$, instead, $\gamma(t)$ monotonically increases upon increasing the time $t$, at least within the range of validity of the perturbative approach used here.
The time at which the kurtosis reaches its maximum
is heuristically expected to be the typical timescale of the probe-field interaction.
We have then investigated the long-time asymptotic behavior of the excess kurtosis $\gamma(t)$, summarized in Table~\ref{tab:long_kurt}. 
It turns out that when the probe is held in a harmonic trap, its dynamics reflects that of the slow modes of the field. 
In contrast, when the probe is free to diffuse, the possible decay of the excess kurtosis upon increasing time is slower than in the presence of the trap and depends on the coupled dynamics of the probe and the field. 

Together with the effective memory kernel of a confined probe~\cite{Basu2022Dynamics}, our predictions concerning the behavior of the excess kurtosis may also shed light on the problem of the dynamics of a probe in a viscoelastic fluid, which has been recently the subject of theoretical and experimental investigations~\cite{Ginot2022Barrier,Ginot2022Recoil}.
At first sight, the coupling to a Gaussian field considered in this work seems too simplistic to be able to provide a quantitative description of the dynamics of the probe in such a complex fluid. However, recent works \cite{Ginot2022Barrier,Ginot2022Recoil} have shown that this dynamics is actually described, quantitatively and at least in some cases, by a model in which the probe is coupled to one or more virtual ``bath particles'' via harmonic springs of stiffnesses $\kappa_i$.
Such a model is linear and therefore the dynamics of the bath particles can be determined analytically and inserted in the equation of motion of the probe, which eventually takes the form of a GLE.
The resulting memory kernel is given by $\Gamma(t)=\sum_i\kappa_i\exp(-\omega_i t)$, where $\omega_i$ is the relaxation rate of the $i$-th bath particle in the potential determined by the spring.

With the model studied here, by contrast, the dynamics of the probe resulting from the coupling to the Gaussian field is non-Markovian and non-linear.
By linearizing this dynamics, one finds a GLE with the same memory kernel as the bath particles model for viscoelasticity, where the sum over the bath particles is replaced by a sum over the Fourier modes of the Gaussian field, as discussed in Sec.~3.2 of Ref.~\cite{Basu2022Dynamics}.
The contribution of the mode $k$ to the effective dynamics of the probe is characterized by the coupling strength $\kappa_k$ and by the relaxation rate $\omega_k$.
Contrary to the bath particles model, the Gaussian field model discussed here encompasses non-linear effects: they arise from the coupling to the mode $k$ if the displacement of the probe exceeds $k^{-1}$ over the timescale $\omega_k^{-1}$ (see Sec.~3.2 of Ref.~\cite{Basu2022Dynamics}).
The Gaussian field model is thus a straightforward generalization of the linear bath particles model, with an additional parameter for each mode, i.e., $k$, which sets the threshold for the appearance of non-linear effects. 
These additional parameters can be used in order to provide a description of the dynamics of a probe in a complex, e.g., viscoelastic, medium in terms of the Gaussian field model studied here (Eqs.~\eqref{eq:energy} and \eqref{eq:dyn_field}) with a suitable choice of the operators $A$, $K$, and $R$.

In particular, one could first adjust the parameters $\kappa_i$ and $\omega_i$ of the bath particles model when the probe dynamics remains linear.
Then, experiments which detect and highlight non-linear effects can be used in order to guide the choice of the additional parameter $k$ of the Gaussian field model for each mode.
Measurements of the two-time correlation of the position of the probe in the trap have revealed the confinement dependence of the effective memory kernel, which is one of such non-linear effects.
Other possible experiments consist in moving the confining potential with a finite velocity~\cite{Venturelli2023Stochastic} or, as suggested here, to measure the kurtosis of the displacement of the probe.

\begin{acknowledgments}
We thank U.~Basu for useful discussions and contributions in the early stages of this work.  A.G.~acknowledges support from MIUR PRIN project ``Coarse-grained description for nonequilibrium systems and transport phenomena (CO-NEST),'' Grant No.~201798CZL.
\end{acknowledgments}

\appendix

\begin{widetext}

\section{Simplification of the cumulant generating function}
\label{app:simp_cgf}

Here we simplify the correction $\psi_2$ to the cumulant generating function, reported in Eq.~\eqref{eq:psiK-fin}.
Changing variable to $s=t'-t''>0$, it reads
\begin{multline}
\psi_2(\qq,t) = -\frac{1}{\gamma}\int \frac{\dd\kk}{(2\pi)^d}\frac{\wt K^2}{\wt A}(\kk\cdot\qq)\int\dd t' \left[\mcR(t-t')-\mcR(-t') \right]\\
\times\int_0^\infty\dd s\,
\ed^{-\wt R\wt A s-\frac{k^2}{2}\sigma_0^2(s)-\kk\cdot\qq\mcS(t,t',t'-s)} \left\{\wt R\wt A+D_0 \left[\kk\cdot\qq[\mcR(t-t'+s)-\mcR(s-t')]+k^2\mcR(s) \right] \right\}.
\end{multline}
Introducing the argument of the exponential (for simplicity we do not write all the arguments of $f(s)$),
\begin{equation}
f(s) = \wt R\wt A s+\frac{k^2}{2}\sigma_0^2(s)+\kk\cdot\qq\,\mcS(t,t',t'-s),
\end{equation}
the derivative of which is
\begin{equation}
f'(s) = \wt R\wt A+D_0 \left\{k^2\mcR(s)+\kk\cdot\qq \left[\mcR(t-t'+s)-\mcR(s-t')+\mcR(t'-s) \right] \right\},
\end{equation}
we can write the correction as
\begin{equation}
\psi_2(\qq,t) = -\frac{1}{\gamma}\int \frac{\dd\kk}{(2\pi)^d}\frac{\wt K^2}{\wt A}(\kk\cdot\qq)\int\dd t' \left[\mcR(t-t')-\mcR(-t') \right]\int_0^\infty\dd s\, 
\ed^{-f(s)} \left[f'(s)-D_0\, \kk\cdot\qq\, \mcR(t'-s)\right].
\end{equation}
Since $f(0)=0$ and $\lim_{s\to\infty}f(s)=\infty$, $\int_0^\infty\ed^{-f(s)}f'(s)\dd s = 1$; then we use that $\int\dd t' \left[\mcR(t-t')-\mcR(-t') \right]=0$.
We arrive at
\begin{align}
\psi_2(\qq,t) & = \frac{D_0^2}{T}\int \frac{\dd\kk}{(2\pi)^d}\frac{\wt K^2}{\wt A}(\kk\cdot\qq)^2\int_0^\infty\dd s \int\dd t' \ed^{-f(s)}\left[\mcR(t-t')-\mcR(-t') \right] \mcR(t'-s)\\
& = \frac{D_0^2}{T}\int \frac{\dd\kk}{(2\pi)^d}\frac{\wt K^2}{\wt A}(\kk\cdot\qq)^2\int_0^t\dd s \int_s^t\dd t' \ed^{-f(s)}\ed^{-\omega_0(t-s)}\\
& = \frac{D_0^2}{T}\ed^{-\omega_0 t}\int \frac{\dd\kk}{(2\pi)^d}\frac{\wt K^2}{\wt A}(\kk\cdot\qq)^2\int_0^t\dd s  \ed^{(\omega_0-\wt R\wt A) s-k^2\sigma_0^2(s)/2}\int_s^t\dd t' \ed^{-\kk\cdot\qq\mcS(t,t',t'-s)}.
\end{align}
In the exponent,
\begin{align}
\mcS(t,t',t'-s)&=\frac{D_0}{\omega_0}\left[\mcR(t-t')-\mcR(t-t'+s)-\mcR(t')+\mcR(t'-s) \right]\\
& = \frac{D_0}{\omega_0} \left[\ed^{-\omega_0(t-t')}-\ed^{-\omega_0(t-t'+s)}-\ed^{-\omega_0 t'}+\ed^{-\omega_0(t'-s)} \right]\\
& = \frac{4D_0}{\omega_0}\ed^{-\omega_0 t/2}\sinh \left(\frac{\omega_0 s}{2} \right)\cosh(\omega_0 u),
\end{align}
where $u=t'-(t+s)/2$.
The integral over $t'$ becomes
\begin{equation}
\int_s^t\dd t' \,\ed^{-\kk\cdot\qq\mcS(t,t',t'-s)}= \frac{2}{\omega_0}J_0 \left(\frac{4D_0}{\omega_0}\, \kk\cdot\qq\, \ed^{-\omega_0 t/2}\sinh \left(\frac{\omega_0 s}{2} \right) ,\frac{\omega_0(t-s)}{2}\right), 
\end{equation}
where $J_0$ is the lower-incomplete form of the modified Bessel function of the second kind reported in Eq.~\eqref{eq:lif_bessel}. 
Finally, by using the invariance of $\psi(\qq,t)$ under spatial rotations of $\qq$, we can write the full correction as in Eq.~\eqref{eq:cgf2}.

\end{widetext}

%

\end{document}